\documentstyle[12pt]{article}

\def\thefootnote{\fnsymbol{footnote}}

\newcommand{\eq}{\begin{equation}}
\newcommand{\en}{\end{equation}}
\newcommand{\eqa}{\begin{eqnarray}}
\newcommand{\ena}{\end{eqnarray}}

\newcommand{\PL}[1]{Phys.\ Lett.\ {\bf #1}}

\newcommand{\PR}[1]{Phys.\ Rev.\ {\bf #1}}
\newcommand{\PRL}[1]{Phys.\ Rev.\ Lett.\ {\bf #1}}

\begin{document}
\begin{titlepage}
\vskip0.5cm
\begin{flushright}
DFTT 66/97\\
HUB-EP-97/96\\
\end{flushright}
\vskip0.5cm
\begin{center}
{\Large\bf The stability of the $O(N)$ invariant }
\vskip 0.3cm
{\Large\bf fixed point in three  dimensions}
\end{center}
\vskip 1.3cm
\centerline{
M. Caselle$^a$\footnote{e--mail: caselle~@to.infn.it}
 and M. Hasenbusch$^b$\footnote{e--mail: hasenbus@birke.physik.hu-berlin.de}}
 \vskip 1.0cm
 \centerline{\sl  $^a$ Dipartimento di Fisica
 Teorica dell'Universit\`a di Torino}
 \centerline{\sl Istituto Nazionale di Fisica Nucleare, Sezione di Torino}
 \centerline{\sl via P.Giuria 1, I-10125 Torino, Italy}
 \vskip .4 cm
 \centerline{\sl $^b$ Humboldt Universit\"at zu Berlin, Institut f\"ur Physik}
 \centerline{\sl Invalidenstr. 110, D-10099 Berlin, Germany}
 \vskip 1.cm

\begin{abstract}
We study the stability of the $O(N)$ fixed point in three dimensions under
perturbations of the cubic type. We address this problem in the three cases
$N=2,3,4$ by using finite size scaling techniques and high precision Monte Carlo
simulations. It is well know that 
there is a critical value $2<N_c<4$ below which
the $O(N)$ fixed point is stable and above which 
the cubic fixed point becomes
the stable one. While we cannot exclude that $N_c<3$, as recently claimed 
by Kleinert and collaborators, our analysis strongly suggests that $N_c$
coincides with 3.

\end{abstract}
\end{titlepage}

\setcounter{footnote}{0}
\def\thefootnote{\arabic{footnote}}

\section{Introduction}

Quantum field theories with $\phi^4$ type interactions are of
importance in several physical contexts. In particular, they represent
one of the most powerful tools in the study of critical 
phenomena~\cite{review}. Due to their
simplicity they allow  perturbative expansions up to rather large orders from
which one can extract  estimates for  various
critical quantities (critical indices and amplitude ratios)  comparable
in precision with those of the most advanced Monte Carlo simulations. In the
simplest case the theory contains a single field $\phi$ and describes the Ising
universality class (for a recent
comparison between field theoretic and Monte Carlo predictions in this case 
see for instance~\cite{ch}).

When the field $\phi$ has more than one component
the situation becomes more complex and different quartic interaction terms 
 can be defined. The simplest one  has the form
$(\sum_{i=1}^N\phi_i^2)^2$. It is $O(N)$ symmetric and describes the $O(N)$
universality class to which belongs, for instance, the isotropic
$N$-component Heisenberg ferromagnet. Besides this term,
the most interesting 
additional contribution is $\sum_{i=1}^N\phi_i^4$ which breaks the $O(N)$
symmetry but
 preserves the cubic invariance. The cubic subgroup of $O(N)$
 is composed of permutations and reflections of the $N$ components of the 
 field.
 Note that in the following "cubic" 
 always refers to the symmetry and not to a third power.
 The importance of the cubic term
 is due to the fact that in a real
crystal the crystalline structure gives rise to
anisotropies which are mainly
of the cubic type. Thus real crystals are better described by mixed actions
in which both the $O(N)$ and the cubic term are present.

Besides this
phenomenological reason, this mixed model is also interesting in itself as it
is a simple non-trivial QFT with different fixed points in
competition among them.
In fact, it is easy to see that in this model there are four possible fixed
points: the trivial Gaussian one, the Ising one (which corresponds to the
situation in which the $N$ components $\phi_i$ decouple), the $O(N)$ symmetric
and the cubic one (see fig.1). It was shown more than twenty years 
ago~\cite{ah73} that while the
Gaussian and Ising fixed points are always unstable, the $O(N)$ and cubic ones
interchange their role as $N$ increases. For $N<N_c$ the $O(N)$ symmetric point
is stable, while for $N>N_c$ it is destabilized by the cubic interaction and the
cubic fixed point becomes the stable one (see fig.1). 
It is possible to see within the
framework of the $\epsilon$-expansion that in three dimensions $N_c<4$. 
The common lore, (supported by $\epsilon$-expansion up to the third order) has 
always been that $N_c$ should lie somewhere between 3 and 4 in three
dimensions, thus implying that the $N=3$ case, which is the most interesting
one for applications to real crystals, should have a stable O(3) symmetric
fixed point. 

In these last years, this commonly
accepted scenario has been contrasted in a series of 
papers~\cite{mss,nas,kt,ks,kst} which suggested that $N_c$ should lie
{\sl below} 3.
As a consequence the critical behaviour of magnetic transitions
in real crystals should be described by the cubic symmetric fixed point, a
result which, if confirmed, would be of relevant interest from
a theoretical point of view.

The aim of this paper is to test this conjecture 
with a high precision Monte Carlo simulation. By studying finite size
corrections of a cubic invariant perturbation term exactly at the critical 
$O(N)$ 
point we can extract the eigenvalues of the stability matrix of the
$O(N)$ fixed point. We study the three interesting cases $N=2,3,4$.
 For $N=2$ and $N=4$
the expected results (stability of the isotropic and cubic fixed point
respectively) are immediately visible from the data. In the $N=3$ case our
 results imply that $N_c \approx 3$.
Obviously  the 
numerical simulation can not decide whether $N_c = 3$ is an exact result.
However we obtain
an upper bound for the absolute value of
the stability index $|b_2|$ for the
O(3) fixed point, 
which turns out to be impressively
small. In particular we are able to exclude all the existing
estimates~\cite{mss,nas,kw,gkw,yh,nr} except
 that of Kleinert and collaborators~\cite{kt,ks,kst}, which is still
 compatible, within one standard deviation, with our result.

\section{The cubic model}

We are interested in the three dimensional quantum field theory defined by 
the Lagrangian
\eq
\label{2.1}
{\cal L}=\frac12\sum_{i=1}^N(\partial_\mu\phi_i \partial^\mu\phi_i +
m^2\phi_i^2) + \frac{\lambda}{4!}(\sum_{i=1}^N\phi_i^2)^2 +
\frac{\mu}{4!}\sum_{i=1}^N\phi_i^4~~~~.
\en

While the term $(\sum_{i=1}^N\phi_i^2)^2$ is  $O(N)$ symmetric,
the term $\sum_{i=1}^N\phi_i^4$ is only invariant under the 
``cubic'' subgroup 
composed by permutations and reflections of the $N$ components $\phi_i$. 
This model is discussed in great detail in several quantum field theory
text books~\cite{review}.

In the following we shall review the first order results in the 
$\epsilon$-expansion.
This rather simple approximation
already gives all the qualitative features of the renormalization flows
of the model.

The fixed points of the theory are given by
the zeros of the $\beta$-functions. 
The stability matrices are given by  
the derivatives of the
$\beta$-functions at the zeros. From the 
eigenvalues of these matrices it is then easy to identify the stable 
fixed point.

The two $\beta$-functions are given, at the first order in the
$\epsilon$-expansion, by

\eq
\label{2.2}
\beta_u=-\epsilon u + u^2\frac{N+8}6 +uv
\en
\eq
\label{2.3}
\beta_v=-\epsilon v + \frac32v^2 + 2uv \;\; ,
\en
where $u$ and $v$ are the renormalized couplings related
to $\lambda$ and $\mu$  respectively.

By looking at the zeros of the $\beta$-functions one sees that
there are four possible fixed points:
\begin{description}
\item{1]}
 The Gaussian fixed point~~~$u=0,~~v=0$  .

\item{2]}
 The Ising fixed point~~~$u=0,~~v=\frac23\epsilon$  .

\item{3]}
The Heisenberg ($O(N)$ invariant) 
fixed point~~~$u=\frac{6}{N+8}\epsilon,~~v=0$  .

\item{4]}
 The Cubic fixed point~~~$u=\frac2N\epsilon,~~v=\frac{2(N-4)}{3N}\epsilon$
.
\end{description}

\noindent
The stability matrix is defined as 
\eq
\label{2.4}                              
B=\left(\matrix{\frac{\partial \beta_u(u,v)}{\partial u}
 & \frac{\partial \beta_u(u,v)}{\partial v}
 \cr\frac{\partial \beta_v(u,v)}{\partial u}
&  \frac{\partial \beta_v(u,v)}{\partial v}}\right)
\;\; .
\en
At the first order of the $\epsilon$-expansion one obtains
\eq
\label{2.5}                              
B=\left(\matrix{-\epsilon + \frac{N+8}{3}u+v
 & u
 \cr 2v
& -\epsilon + 2u+3v }\right)
\;\; .
\en
 The corresponding eigenvalues, evaluated at the four fixed points are:

\begin{description}
\item{1]}
 Gaussian :~~~$b_1=b_2=-\epsilon$   .

\item{2]}
 Ising :~~~$b_1=-\frac\epsilon3,~~b_2=\epsilon$  . 

\item{3]}
 Heisenberg :~~~~$b_1=\epsilon,~~b_2=\frac{4-N}{N+8}\epsilon$  .

\item{4]}
 Cubic fixed point:~~~~$b_1=N\epsilon,~~b_2=\frac{N-4}{3}\epsilon$  .

\end{description}
It is easy to see that the Gaussian and Ising fixed points are 
always unstable,
independently from the value of $N$. 
In particular the Ising f.p. has only one
direction of instability, while the Gaussian one is unstable 
in both directions.

The cubic and $O(N)$ fixed points interchange 
their role as a function of $N$.
For $N$ smaller than a critical value $N_c$ (which at this order in the 
$\epsilon$-expansion turns out to be 4) 
the Heisenberg fixed point is the stable one and defines the
universality class
toward which the system flows in the infrared limit.

For $N>N_c$, $b_2$ evaluated at the Heisenberg point
becomes negative while
$b_2$ evaluated at the cubic f.p. becomes positive
and the cubic fixed point
becomes the stable one.

The renormalization flows corresponding to these two situations are
reported in fig.1. For $N<N_c$ all initial points with $u>0$ and 
$v>\frac{N-4}3u$ will flow toward the $O(N)$-invariant,
Heisenberg fixed point. 
For $N>N_c$ all initial points with $u>0$ and $v>0$ 
will flow in the infrared limit toward the cubic fixed point which, for 
$N<N_c$ lies in the $v<0$ half plane, exactly at
$N=N_c$ crosses the $v=0$ axis and moves for $N>N_c$ in the $v>0$ region. 
 
Initial points outside the above defined regions
flow away toward more negative
values of $u$ and/or $v$ and finally reach
the region in which the positivity
condition for the quartic potential is no longer satisfied.
These trajectories 
are related (from the statistical mechanics point of view)
to realizations of
the cubic model in which the phase transition
is of the fluctuation-induced 
first order type.
These models have recently attracted much interest
as a laboratory to study
arbitrary weak first order transitions~\cite{nt}.

The last remaining point is now to find the value of $N_c$ in three 
dimensions. It is easy to see
by looking to higher orders in the $\epsilon$-expansion, or with the help 
of Monte Carlo simulations, that  for $N=2$ the Heisenberg fixed point
is stable
and that on the contrary  for $N=4$ the cubic fixed point is the 
stable one.
Thus $2<N_c<4$. However it turns out to be hard to decide whether $N_c$ is
greater or lower than 3.
Equivalently one can look at the sign of the $b_2$ 
eigenvalue at the Heisenberg point for $N=3$. 
If $b_2$ is positive, then $N_c$
must be greater than 3.
In the past twenty years much efforts have been devoted to
settle this question. The first result was reported in~\cite{kw} where  
the $\epsilon$-expansion for $N_c$ was extended up to the third order
leading to the estimate
$N_c=3.128$.
In agreement with this estimate (but,  using a completely different
approach), Grover, Kadanoff and Wegner~\cite{gkw} obtained $b_2=0.053$
at $N=3$. Few years later, with different approximation techniques,
 the two contrasting results: $N_c\sim 2.3$~\cite{yh} and 
$N_c\sim 3.4$~\cite{nr} were obtained. Ten years later in~\cite{mss,nas}
a value $N_c<3$ was suggested. In particular in~\cite{nas}, by means of a
three loop calculation directly in $d=3$, the values $N_c=2.91$ and $b_2=-0.
008$ were proposed. Finally, more
recently, Kleinert and collaborators pushed
the $\epsilon$-expansion up to the fifth order
\cite{ks} and obtained a similar answer. First,
in~\cite{ks} they found,
(with a [2,2] Pade' approximant) $N_c=2.958$.
Then in~\cite{kst}, by using a careful resummation procedure of the
fifth order series, they obtained the value $b_2=-0.00214$
for the stability eigenvalue at $N=3$. Due to the nature
of these results it is very difficult to add sensible error-bars to these
estimates. However, it is clear from the above discussion that the existing
estimates for $N_c$ are scattered around $N_c=3$ and that as  the
various techniques and approximations become more and more refined the 
corresponding estimates for $N_c$ get closer and closer to $N_c=3$.

\section{The simulation}

\subsection{The model}
The cubic model discussed in sect.2 has a simple and straightforward
lattice realization, defined by the action:
\begin{equation}
\label{3.1}
S= - \beta \sum_{<xy>} s_x s_y  - \mu \sum_x \sum_{i=1}^N (s_x^i)^4 \;\; ,
\end{equation}
where $s_x$ is a unit vector in $R^N$.  
$<x,y>$  denotes a pair of nearest neighbour sites on the lattice.
We consider a three-dimensional cubic lattice of size $L$ and lattice
spacing $a=1$.
For $\mu=0$ we have the 
standard $O(N)$ invariant (Heisenberg) model, while for $\mu\neq0$ the 
cubic-invariant 
perturbation $\sum_x \sum_{i=1}^N (s_x^i)^4$ breaks the $O(N)$ symmetry. 
In the following we shall study this model in the three
cases $N=2,3,4$. 
We shall concentrate our main efforts in the $N=3$ case.

In three dimensions, for $\mu=0$, the $O(N)$ model 
undergoes a second order phase transition for
some value $\beta_c$ (which depends on $N$) of the coupling. In
the vicinity of such a point  the continuum limit can be taken, leading
to the  $O(N)$ symmetric QFT (corresponding to the $v=0$ axis in fig.1) 
discussed in the previous section. The presence of such a
continuous phase transition is obviously a mandatory 
condition for the whole analysis.

The simplest way to extract the value of $N_c$ from a lattice simulation 
is to determine the stability eigenvalues  $b_2$ for $N=3$.
  From the lattice point of
view the $b_2$ eigenvalue appears as the critical 
index which controls the behaviour of any cubic-invariant 
(but $O(N)$-violating) observable in the vicinity of the 
Heisenberg transition point.

The most efficient way to evaluate such a critical index is to look at the
finite size dependence (as a function of the lattice size $L$) of a suitable
observable (to be defined below) evaluated exactly at the critical point
$\beta_c$. To this end it is necessary to have a very good estimate of 
the critical coupling.
Fortunately, $\beta_c$ is known with very high precision
in each of the three cases $N=2,3,4$ in which we are interested. 
This is one of
the reasons for which we have chosen this particular lattice realization of the cubic model.

In tab.1,2,3 we have collected
 the most recent results for $\beta_c$
 both from Monte Carlo simulations 
 and from series expansions. HT-$\theta$ indicates the
 biased resummation of the HT series in which the value of the index 
$\theta$ is given as input parameter. It is interesting to see that all the
estimates agree within the errors.

In tab.4 we report the values (chosen from tab.1,2,3) that we used in our 
simulations.

\begin{table}[h]
\label{literature1}
\caption{\sl Results for $\beta_c$  given in the literature for $N=2$}
\vskip 0.2cm
\begin{tabular}{|c|c|c|}
\hline
ref. &method & $\beta_c$ \\
\hline
\cite{wir1}& MCRG & 0.45420(2)  \\
\cite{wir1}& MC & 0.454170(7)  \\
\cite{bfmm}& MC & 0.454165(4)  \\
\cite{bc}& HT & 0.45419(3)  \\
\hline
\end{tabular}
\end{table}

\begin{table}[h]
\label{literature2}
\caption{\sl Results for $\beta_c$  given in the literature for $N=3$}
\vskip 0.2cm
\begin{tabular}{|c|c|c|}
\hline
ref. &method & $\beta_c$ \\
\hline
\cite{Janke2}& MC & 0.6930(1)  \\
\cite{bc2}& MC & 0.6931(1)  \\
\cite{cfl}& MC & 0.693035(37)  \\
\cite{bfmm}& MC & 0.693002(12)  \\
\cite{bc}& HT & 0.69303(3)  \\
\cite{bc}& HT-$\theta$ & 0.69305(4)  \\
\hline
\end{tabular}
\end{table}

\begin{table}[h]
\label{literature3}
\caption{\sl Results for $\beta_c$  given in the literature for $N=4$}
\vskip 0.2cm
\begin{tabular}{|c|c|c|}
\hline
ref. &method & $\beta_c$ \\
\hline
\cite{kk}& MC & 0.9360(1)  \\
\cite{bfmm}& MC & 0.935861(8)  \\
\cite{bc}& HT & 0.93589(6)  \\
\cite{bc}& HT-$\theta$ & 0.93593(6)  \\
\hline
\end{tabular}
\end{table}

\begin{table}[h]
\label{betac}
\caption{\sl Values of $\beta_c$  used in this article.}
\vskip 0.2cm
\begin{tabular}{|c|c|}
\hline
N & $\beta_c$ \\
\hline
 2 & 0.454165(4)  \\
 3 & 0.693002(12)  \\
 4 & 0.935861(8)  \\
\hline
\end{tabular}
\end{table}

\subsection{Observables}
There are four natural observables in the model (\ref{3.1}).
The two terms which appear in the action:
\begin{equation}
E\equiv \sum_{<xy>} s_x s_y
\end{equation}
and 
\begin{equation}
P\equiv \sum_x \sum_i  (s_x^i)^4 \;\; .
\end{equation}

The total magnetization, which is the order parameter of the transition: 
\begin{equation}
M\equiv \sum_x s_x
\end{equation}
 and the ratio
\begin{equation}
R= \frac{ \sum_i (M^i)^4 }{(M^2)^2}  
\end{equation}
which quantifies the violation of the $O(N)$ symmetry in the model
\footnote{Other choices are possible for 
this last observable. For instance the term
\begin{equation}
X= \frac{M_{max}^2}{M^2}  \;\; ,
\end{equation}
where $M_{max}^2$ is the maximal square of a component of the 
magnetization would work equally well. However it turns out that the ratio 
$R$ defined above is the one
which can be measured in the most efficient and simple way.}.

In order to study the stability of the fixed point we are actually interested
in  the 
derivative of $<R>$  with respect to $\mu$ at $\beta=\beta_c$
and $\mu=0$: 
\begin{equation}
D_{R} \equiv \frac{\partial <R>}{\partial \mu} 
\vert_{\mu=0,\beta=\beta_c}  \;\; .
\end{equation}

In fact this derivative measures the 
(generalized) susceptibility of the system
with respect to a cubic perturbation,
just like what happens for the ordinary 
magnetic susceptibility in the case of a magnetic perturbation or for the 
specific heat in the case of a thermal perturbation. 
At this point, a standard finite size scaling analysis tells us
that the critical index which measures the infrared stability
of the system with respect to the above perturbations, also controls 
the finite size behaviour (namely the $L$ dependence) of the corresponding
susceptibility exactly at the critical point.
In particular, in the case in which we are interested, $D_R$ should behave 
for $\beta=\beta_c$ and $\mu=0$, as

\begin{equation}
\label{scaling}
 D_R \propto  L^{-b_2} \;\; .
\end{equation}

where $b_2$  is exactly the stability 
eigenvalue of the Heisenberg fixed point
that we are looking for.

Scaling laws of the type (\ref{scaling}) are expected to hold
for sufficiently large values of L.  
For small lattices,
correction to scaling terms should be expected. It is thus 
important to have results with small 
statistical errors for large lattice sizes.
Few preliminary tests have shown us that (at least for the
 case $N=3$) sizes up to $L=32$ are needed  to extract reliable 
estimates of $b_2$. This
requirement represents the major technical problem of this work.

There are two possible choices to compute  $D_R$:
\begin{itemize}
\item
It can be computed directly in the simulation at $\mu=0$ as
\begin{equation}
D_{R} = \langle P R \rangle - 
 \langle P \rangle \langle R \rangle  \;\; .
\end{equation}
\item
It can be computed by using the finite difference method, i.e. by simulating the
model at small non-zero values of $\mu$:
\begin{equation}
\label{dmu}
D_{R}(\mu) \equiv  \frac{R(\mu) - R(-\mu)}{2 \mu} \;\; .
\end{equation}

\end{itemize}
It is easy to recover the relationship between $D_R$ and $D_R(\mu)$. First, let
us notice that
on a finite lattice $R(\mu,\beta)$ must be an analytic function 
of its parameters. This holds  also at $\beta=\beta_c$. Therefore we can
Taylor-expand $R(\mu,\beta)$ in powers of $\mu$ for fixed $\beta=\beta_c$.
Taking the symmetric difference we obtain
\begin{equation}
\label{u1}
D_{R} = D_R(\mu)
  +  \frac1{3!} \frac{\mbox{d}^3R}{\mbox{d}\mu^3} \mu^2 + O(\mu^4) \;\; .
\end{equation}

Both these definitions have their drawbacks.
The $\mu=0$ simulations are affected by 
a strong enhancement of the variance
(hence of the statistical errors) as $L$ increases. It turns out that
the statistical error of $D_R$ at a fixed number of measurements
increases roughly as $L^{3/2}$. As a consequence too large samples are
needed to keep the error sufficiently small for $L>16$.

On the contrary for $D_{R}(\mu)$ at fixed $\mu$ the statistical error
does not increase with $L$. However the $O(\mu^2)$
corrections do increase with $L$. Reducing these $O(\mu^2)$
corrections requires to reduce the value of $\mu$. This in turn 
requires to increase the 
number of measurements to keep the statistical error fixed.

In the following section we shall discuss a way out of these problems. 
By using  the global $O(N)$ symmetry of the model  at $\mu=0$
an improved version of $D_R$ can be constructed.
 This improvement does not change the $L$ dependence of the variance,
but gives a significant reduction of its magnitude, 
thus allowing to reach, with
a reasonable CPU time, lattices sizes as large as $L=32$ 
which are large enough to
extract the finite size behaviour with the required precision.   
Most of the data that we shall
discuss in the last section have been obtained
by using this improved observable. We also performed, as a cross check,
some simulations at finite $\mu$. The agreement that we find between
the values of $D_R$ obtained in these two ways 
is a non trivial check of the reliability of our results.

\subsection{Variance reduced estimator for $D_R$.}
Variance reduced estimators have the same expectation value as the 
corresponding standard estimators. However their variance is reduced, 
which allows for more accurate results in Monte Carlo simulations 
than the standard estimator.
A general principle to construct variance reduced estimators is to 
look for degrees of freedom which can be integrated out analytically. 

In order to obtain a variance reduced estimator of
$\frac{\partial <R>}{\partial \mu}$ 
we integrate $P$,$R$ and $ P R $  
over the global $O(N)$ rotations.


This is trivial for  $P$ and 
 $R$, but it is less simple in the case of the $PR$ component. Since 
 $P$ is a sum over all lattice sites we can commute the integration 
 over the global rotations  and the summation over the lattice sites.
 The integral to be solved is hence given by
\begin{equation}
\label{int}
I = \int_{O(N)} \mbox{D}T \left(\sum_i [(T s_x)^i]^4 \right) 
 \left(\sum_i [(T m)^i]^4 \right) \;\; ,
 \end{equation}
 where $T$ is an element of $O(N)$, $\mbox{D}T$ the Haar-measure, 
 $s_x$ the spin at the site $x$, and $m$ a unit vector in the direction
 of the global magnetization. For symmetry reasons the integral only
 depends on the angle between $s_x$ and $m$, which we define as follows: 
\begin{equation}
m s_x = \cos(\alpha)~~~~.
\end{equation}

The integral (\ref{int}) can be evaluated explicitly for any value of $N$.
Details of the calculation are reported in the appendix. Here we only list 
the results in the three cases in which we are interested:

\begin{description}
\item{$N=2$}
\begin{equation}
I = \frac{9}{16} + \frac{1}{32}\cos(4 \alpha)
 \end{equation}

\item{$N=3$}

\eq
I=\frac{1}{60} \cos(4\alpha) + \frac{1}{105} \cos(2\alpha) + \frac{153}{420}
\en

\item{$N=4$}

\eq
  I = \frac{2}{25} \cos^4(\alpha) - \frac{3}{50} \cos^2(\alpha) + 
\frac{51}{200}
\en

\end{description}

\subsection{The Monte Carlo Algorithm}
Due to the different symmetries in the models, we had to use different 
algorithms in the two cases $\mu=0$ and $\mu\neq0$.

\subsubsection{$\mu=0$}
In the case $\mu=0$ we used the single cluster algorithm of U.Wolff
\cite{ulli} and the microcanonical overrelaxation algorithm. 
The basic idea of the cluster-algorithm
is to construct conditional Ising models.
This is achieved by allowing only the sign-change of the spin component 
parallel to an unit vector $r$ in $R^N$.  
The delete probability depends on the pair of lattice sites and is 
given by
\begin{equation}
p_d(x,y) = \mbox{min} [1,\exp(-2 (r s_x)(r s_y))]  \;\; ,
\end{equation}
where $x$ and $y$ are nearest neighbour sites on the lattice.
The vector $r$ is chosen with a probability density
uniform on $S^{N-1}$.  For each update a new $r$ is chosen.

The variance of the
improved estimator of $D_R$ is mainly caused by local fluctuations of 
the spins. Hence it is useful to supplement the cluster-algorithm 
with a fast local algorithm to produce local changes of the configuration.
For that purpose we used the microcanonical overrelaxation algorithm.

The elementary update of the algorithm is given by 
\begin{equation}
 s_x' = \frac{2 \; (n_x s_x) \; n_x}{n_x^2} -s_x   \;\; ,
\end{equation}
where $n_x$ is the sum of the nearest neighbour spins of $s_x$.
Since neither a random number nor the evaluation of the exponential
function is needed for this update the CPU-time required is rather 
small compared with a Metropolis or a heat-bath update.

The whole update cycle that we used in our simulations 
consists of  a mixture of
9 overrelaxation sweeps and $K$ cluster-updates .
$K$ is chosen such that the number of spins updated in 
$K$ cluster-updates is of the order of the number of lattice-sites.
A measurement 
is performed after each overrelaxation sweeps and after the 
cluster-updates. There are hence 10 measurements in the 
9 overrelaxation sweeps and $K$ cluster-updates cycle.

\subsubsection{$\mu\neq0$}
For $\mu\ne0$ some modifications are needed.  
We have restricted the vector $r$ such 
that $\sum_i (s^i)^4$ is not changed by the update. This 
is guaranteed if the sign of a component is changed or 
two components are exchanged or a combination of both. 
This means that $r$ is either parallel
to an axis or is diagonal in a plane.

$r$ is in
\begin{equation}
(1,0,...,0) , (0,1,...,0) ... (0,0,...,1)
\end{equation}

or 

\begin{equation}
\frac{1}{\sqrt 2} (1,1,...,0) , 
\frac{1}{\sqrt 2} (1,0,...,1) ,
... \frac{1}{\sqrt 2} (0,...1,1)
\end{equation} 
or
\begin{equation}
\frac{1}{\sqrt 2} (1,-1,...,0) , 
\frac{1}{\sqrt 2} (1,0,...,-1) ,
... \frac{1}{\sqrt 2} (0,...1,-1)   \;\; .
\end{equation}

While this restriction of $r$ to a discrete subset of $S^{N-1}$ 
does not violate detailed balance it means that the cluster-update 
by itself is not ergodic.

In order to restore ergodicity we supplement  the cluster-update 
with an (ergodic) Metropolis update. For performance reasons we also
added  a local reflection 
update that is microcanonical for $\mu=0$. A spin 
is reflected at the sum of its neighbours.
\begin{equation}
 s_x' = \frac{(S_x s_x) S_x}{S_x^2} - 2 s_x \;\; .
\end{equation}
where $S_x$ is the sum of the spins on nearest neighbour sites of $x$.
For $\mu\ne0$  the proposal $s_x'$ is accepted with a probability 
\begin{equation}
P_{acc} =\mbox{min}[1,\exp(\mu \sum_i [(s_x'^i)^4 - (s_x^i)^4])] \;\;.
\end{equation}

A whole update cycle consists of one Metropolis update,
one local reflection update plus $K$ cluster-updates.
$K$ is chosen, as in the $\mu=0$, case such that the number of spins 
updated in 
$K$ cluster-updates is of the order of the number of lattice-sites.
A measure is performed in each update cycle.

\subsection{Statistical and systematic errors}
We evaluated statistical errors with the standard binning method. 
Both in the $\mu=0$ and $\mu\neq0$ cases bins of 
1.000 update cycles are chosen
(corresponding to 10.000 and 1.000 measurements  respectively).
This binning was already performed during the simulation since not all 
individual measurements could be stored on disc.
Besides the
statistical uncertainty we have to face also 
the systematic error due to the
uncertainty $\Delta \beta_c$ in the estimate of $\beta_c$. To evaluate this
error we also measured in the simulation  the expectation value 
$<E R>$. The difference
\eq
 <E R> -<E><R>
\en
gives an estimate of the derivative of $<R>$ with respect to $\beta$, 
from which we can obtain the systematic error induced on $R$ by the
uncertainty in $\beta_c$: 
\eq
(<E R> -<E><R>)*\Delta \beta_c \;\; .  
\en

With a similar construction we obtain the error induced on $D_R$.
We can consider this as a lower bound on the accuracy 
that we can reach for the derivatives.    
It makes no sense to reduce the statistical error of $D_R$ below
this bound. This observation fixes the typical sample size for the simulations,
which turned out to be of the order of 20.000 bins.

\section{Results and discussion}

\subsection{Results at $\mu=0$}

We simulated the models with $N=2,3,4$ at $\mu=0,~~\beta=\beta_c$, in the
ranges $L\in[4-16]$ for $N=2$; $L\in[4-32]$ for $N=3$ and 
$L\in[4-20]$ for $N=4$.
The results are reported in  tables 5, 6 and 7. In the first column we
report the lattice size and in the second the values of the derivative $D_R$.
the first error in parenthesis denotes the statistical uncertainty, while in the
second parenthesis the error induced by the uncertainty in
$\beta_c$  is reported. In the last column we report
the sample size (number of bins times number of measurements in each bin). 

\begin{table}[h]
\label{resn=2}
\caption{\sl Results for $D_R$  in the  $N=2$ model}
\vskip 0.2cm
\begin{tabular}{|c|c|c|}
\hline
  L &   $D_R$ & statistics \\
\hline
  4 &  .011046(20)(1)         &    $10000*10^4$ \\
  6 &  .010506(34)(2)         &    $10000*10^4$ \\
  8 &  .010121(51)(2)         &    $10000*10^4$ \\
 10 &  .009728(58)(4)         &    $15000*10^4$ \\
 12 &  .009523(76)(5)         &    $15000*10^4$ \\
 16 &  .008904(117)(4)        &    $15000*10^4$ \\
\hline
\end{tabular}
\end{table}

\begin{table}[h]
\label{resn=3}
\caption{\sl Results for $D_R$  in the  $N=3$ model}
\vskip 0.2cm
\begin{tabular}{|c|c|c|}
\hline
  L &   $D_R$ & statistics \\
\hline
  4 &  .019672(19)(3)         &    $10000*10^4$ \\
  6 &  .020118(19)(5)         &    $15000*10^4$ \\
  8 &  .020187(20)(7)         &    $20004*10^4$ \\
 10 &  .020196(25)(10)         &    $20870*10^4$ \\
 12 &  .020152(32)(13)         &    $21500*10^4$ \\
 14 &  .020178(40)(15)         &    $20770*10^4$ \\
 16 &  .020233(49)(16)        &    $20750*10^4$ \\
 20 &  .020094(68)(22)        &    $20150*10^4$ \\
 24 &  .020178(84)(28)        &    $23145*10^4$ \\
 32 &  .020265(140)(39)        &    $19560*10^4$ \\
\hline
\end{tabular}
\end{table}

\begin{table}[h]
\label{resn=4}
\caption{\sl Results for $D_R$  in the  $N=4$ model}
\vskip 0.2cm
\begin{tabular}{|c|c|c|}
\hline
  L &   $D_R$ & statistics \\
\hline
  4 &  .023474(15)(1)         &    $10000*10^4$ \\
  6 &  .025287(16)(3)         &    $10000*10^4$ \\
  8 &  .026294(19)(4)         &    $10000*10^4$ \\
 10 &  .027097(23)(6)         &    $10000*10^4$ \\
 12 &  .027742(27)(8)         &    $10600*10^4$ \\
 16 &  .028774(39)(13)        &    $10000*10^4$ \\
 20 &  .029605(52)(38)        &    $10050*10^4$ \\
\hline
\end{tabular}
\end{table}

\subsection{Results at $\mu\neq 0$}

As a test of the above results we also performed some simulations at 
$\mu\neq0$, both for $N=2$ and $N=3$.
 We evaluated the finite $\mu$ estimators
$D_R(\mu)$ by using eq.(\ref{dmu}).
The results are reported in tab.8 where, in
the last line, we also reported for comparison the corresponding $\mu=0$ 
estimates.

\begin{table}[h]
\label{resmu}
\caption{\sl Results for $D_R$ for $\mu\neq 0$}
\vskip 0.2cm
\begin{tabular}{|c|c|c|c|c|}
\hline
 &$N=2,~L=8$ & $N=2,~L=12$ & $N=2,~L=16$ & $N=3,~L=12$ \\
\hline
 $\mu=4$ & & .010503(140)(3) & & \\
 $\mu=2$ & .010117(28)(1) &  .010008(122)(3) & & \\
 $\mu=1$ &  .010112(57)(1)& .009495(118)(3) & & .020977(64)(9) \\
 $\mu=0.5$ &.010054(74)(1) &.009505(103)(3)  &  .009139(103)(4)&
 .020306(130)(9) \\
 $\mu=0.25$ & &  & &.020076(257)(9) \\
\hline
 $\mu=0$ & .010121(51)(2) & .009523(76)(5) & .008904(117)(4) & 
.020152(32)(13)\\
\hline
\end{tabular}
\end{table}

The agreement between the results obtained with the two approaches is very 
good and makes us confident on the reliability of the $\mu=0$ set of data. 

\subsection{The $b_2$ index}

We fitted the data obtained at $\mu=0$ with the scaling law

\begin{equation}
\label{scal}
 D_R = C~L^{-b_2} \;\; .
\end{equation}
The fit results are collected
in tab.9 where in the second column we give
the minimum value $L_{min}$ of $L$ taken into account in the fit.
In the third and fourth column we report the reduced
$\chi^2$ and the confidence level respectively.
Finally, in the last two column the
best fit values of $C$ and $b_2$ are reported.
As usual we give in the first parenthesis the statistical
error and in the second the error induced by $\beta_c$.
The various fits are plotted and compared in fig.2 and 3.

The large value of $\chi^2$ clearly indicates that 
for any value of $N$ the sample at $L=4$
is strongly affected by correction to scaling 
terms and must be discarded.  Fits without $L=4$ have an acceptable
$\chi^2$. However this fact does not necessarily imply that it is 
justified to ignore corrections to scaling. Hence we regard the fits 
with $L_{min}=8$ as our final result. Still it remains difficult to 
quantify the systematic error due to corrections to scaling. Based
on the experience with the finite size scaling analysis of other exponents  
of the Heisenberg model we expect them to be of the same order of magnitude
as the statistical error of the  $L_{min}=8$ fits.

\begin{table}[h]
\label{resfit}
\caption{\sl Results for $C$ and $b_2$}
\vskip 0.2cm
\begin{tabular}{|c|c|c|c|c|c|}
\hline
$N$ & $L_{min}$ & $\chi^2_{red}$ & C.L.&
 $C$ & $b_2$   \\
\hline
2 &  4 & 2.01 & 9\% & 0.01335(9)(1) &  0.1362(40)(3)  \\
2 &  6 & 1.17 & 32\% & 0.01381(24)(1) &  0.1519(84)(4)  \\
2 &  8 & 0.85 & 43\% & 0.01445(54)(2) &  0.1711(166)(6)  \\
\hline
3 &  4 & 26.5 & 0\% & 0.01936(4)(2) & -0.0174(10)(6)  \\
3 &  6 & 1.32 & 23\% & 0.02005(6)(2) & -0.0026(14)(7)  \\
3 &  8 & 0.71 & 64\% & 0.02022(10)(3) & 0.0007(20)(9)  \\
\hline
4 &  4 & 77.5 & 0\% & 0.01928(3)(1) & -0.1473(7)(4)  \\
4 &  6 & 1.22 & 30\% & 0.01997(5)(2) &-0.1321(10)(5)  \\
4 &  8 & 0.51 & 67\% & 0.02008(8)(3) &-0.1299(16)(8)  \\
\hline
\end{tabular}
\end{table}

\subsection{Discussion and comparison with other estimates}

As it can be seen from tab.9, our results for $N=3$ are certainly incompatible 
with all the existing estimates~\cite{mss,nas,kw,gkw,yh,nr}, 
except that of Kleinert and
collaborators~\cite{kt,ks,kst}. As a matter of fact, 
if we keep  in the fit for $N=3$
also the $L=6$ sample we find an impressive agreement with the result 
$b_2=-0.00214$ of ref.~\cite{ks}. 
However, as mentioned above, we strongly suspect
that the $L=6$ sample is still affected by correction to scaling terms and
prefer to quote as our best estimate the $L=[8-32]$ result
$b_2=0.0007(20)(9)$,
which is still compatible with the result of ref.~\cite{ks}, but 
suggests that $N_c$ could indeed exactly coincide with $3$.
In this respect it must also be noticed that the trend of the perturbative 
estimates of $b_2$ quoted in~\cite{ks} as a function of the order in the 
perturbative expansion also suggests that $N_c$ converges to 3 in agreement
with our result. 

In any case,
let us stress again that it is obviously impossible to decide by means
of a numerical simulation if $N_c=3$ is an exact result and that
the fact that the difference $|N_c-3|$ is so small, and compatible with zero,
 might well be a coincidence.
However we think that it would be worthwhile to look for 
an argument which explains why the cubic and Heisenberg
 fixed point in three dimensions should coincide exactly for N=3.

\vskip 2cm

\vskip 1cm
{\bf  Acknowledgements}

We thank F.Gliozzi, G.M\"unster, 
K.Pinn, P.Provero and S.Vinti for many helpful 
discussions. 
Work supported in part by the European Commission TMR programme 
ERBFMRX-CT96-0045.

\newpage
\appendix{\Large{\bf{Appendix}}}

\vskip 0.5cm

\renewcommand{\theequation}{A.\arabic{equation}}
\setcounter{equation}{0}

In this appendix we evaluate the integral

\begin{equation}
I = \int_{O(N)} \mbox{D}T \left(\sum_i [(T s_x)^i]^4 \right) 
 \left(\sum_i [(T m)^i]^4 \right) \;\; ,
 \end{equation}
 where $T$ is an element of $O(N)$, $\mbox{D}T$ the Haar-measure, 
 $s_x$ the spin at the site $x$  and $m$ a unit vector in the direction
 of the global magnetization. For symmetry reasons the integral only
 depends on the angle between $s_x$ and $m$ which is defined as
\begin{equation}
m s_x = \cos(\alpha) \;\; .
\end{equation}

We write the integral as
\begin{equation}
I(\alpha)=\int_{S^{N-1}} \mbox{d}s \; (\sum_i s_i^4)  
\int_{S^{N-2},s} \mbox{d}t \; (\sum_i t_i^4) \;\; ,
\end{equation}
 where $\int_{S^{N-1}}$ denotes the integral over the $N$ dimensional 
 sphere. $\int_{S^{N-2},s}$ denotes the integral over the $N-1$ dimensional
 subspace defined as the set of all the vectors $t$ that for any fixed $s$
 satisfy the equation
 $st =\cos(\alpha)$.
We choose the normalizations so that $\int_{S^{N-1}} \mbox{d}s =1 $ and 
 $\int_{S^{N-2},s} \mbox{d}t =1$

 Because of symmetry we can restrict the calculation to the 
 first component of $s$
\begin{equation}
I(\alpha)=N \;\; \int_{S^{N-1}} \mbox{d}s \;  s_1^4  
\int_{S^{N-2},s} \mbox{d}t \; (\sum_i t_i^4) \;\; .
\end{equation}

Now we decompose  the integral $\int_{S^{N-1}}$ into the integral
over the $s_1$ component and for fixed $s_1$ over the remaining 
$S^{N-2}$.

We get
\begin{equation}
I(\alpha)=N \;\; const \int_{s_1=0}^{s_1=1}
\mbox{d}s_1 \; (1-s_1^2)^{(N-3)/2} \; s_1^4
\int_{S^{N-2}} \mbox{d}s' \int_{S^{N-2},s} \mbox{d}t \; (\sum_i t_i^4) 
\;\; ,
\end{equation}

where 
\begin{equation}
const_1 
= \left[\int_{s_1=0}^{s_1=1}\mbox{d}s_1 \; (1-s_1^2)^{(N-3)/2}\right]^{-1}
\end{equation}
and $s'$ is $s$ without the 1-component.

Let us now study
\begin{equation}
 \int_{S^{N-2}} \mbox{d}s' \int_{S^{N-2},s}  \mbox{d}t  \;\; .
\end{equation}
This measure for $t$ is invariant under rotations around the 1-axis.
The non-trivial question is the measure for the 1-component of $t$.
The range of $t_1$ is given by
\begin{equation}
t_{max} = \cos(\alpha) s_1 + \sin(\alpha) \sqrt{1-s_1^2}
\end{equation}
and
\begin{equation}
t_{min} = \cos(\alpha) s_1 - \sin(\alpha) \sqrt{1-s_1^2} \;\; .
\end{equation}
The measure between this extreme values is given by the  fact that
for any $s$, $t$ is distributed on a $S^{N-2}$ sphere.
Hence the measure is  (for $N>2$)
\begin{equation}
const_2 \left[1-\left(\frac{t_1-c}{2 s}\right)^2 \right]^{(N-4)/2}
\end{equation}
with $c=\cos(\alpha) s_1$ and $s=\sin(\alpha) \sqrt{1-s_1^2}$ .
The normalization
$const_2$ is given by
\eq
const_2=\left\{ 
 \int_{t_1=c-s}^{c+s} 
  \left[1-\left(\frac{t_1-c}{2 s}\right)^2 \right]^{(N-4)/2}\right\}^{-1}
  \;\; .
\en
For fixed $t_1$ the integration of the remaining  components 
gives us 

\noindent
$ (1-t_1^2)^2 \;  <R>_{N-1} $, with $<R>_N = 3/(N+2)$.

Now we are in the position to write down the full integral:
\begin{eqnarray}
 I(\alpha) &=& N \;\; const_1 \int_{s_1=0}^{s_1=1}
 \mbox{d}s_1 (1-s_1^2)^{(N-3)/2} s_1^4  \\
 &const_2& 
 \int_{t_1=c-s}^{c+s} 
  \left[1-\left(\frac{t_1-c}{2 s}\right)^2 \right]^{(N-4)/2}
  (t_1^4 + \frac{3}{N+1} (1-t_1^2)^2) \;\; .
\end{eqnarray}

This integral can be solved with standard techniques and yields in the three cases
$N=2,3,4$ the results listed in sect. 3

\newpage
{\bf Figure Captions}
\vskip 1cm
\begin{description}
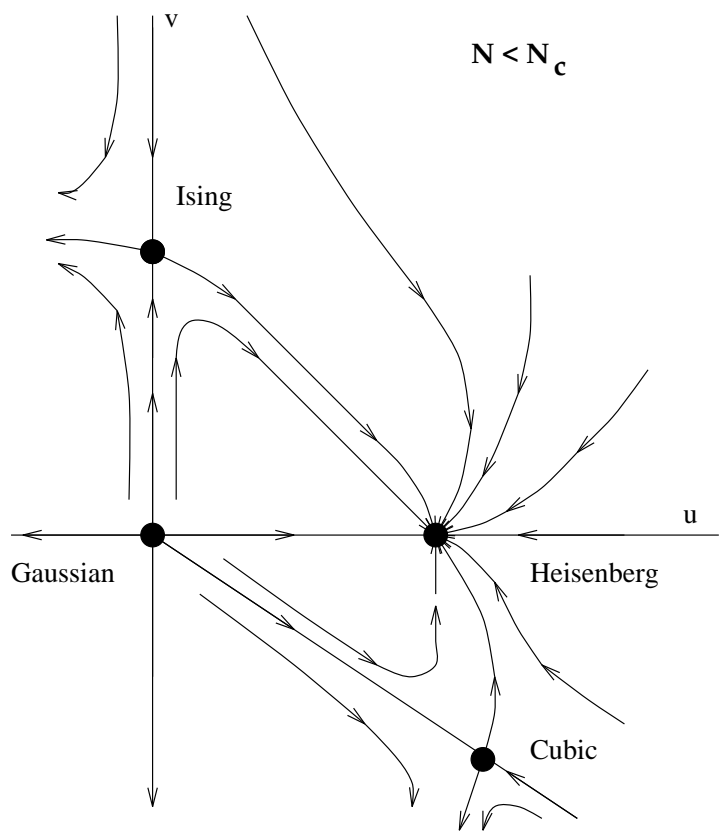
\item{\bf Fig.1} 
Renormalization group flows for the cubic model in three dimensions.
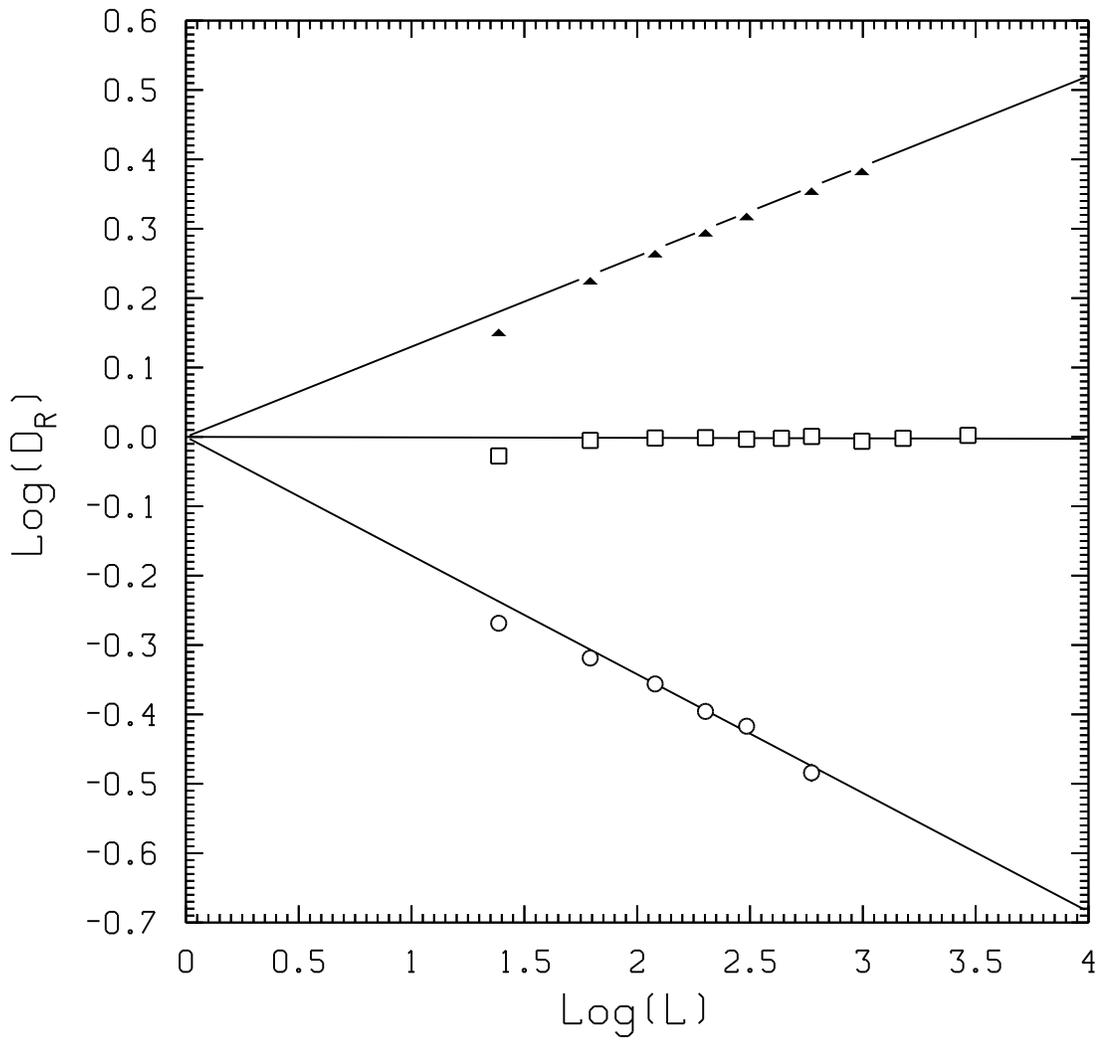
\item{\bf Fig.2}
$Log(D_R)$ as a function of $Log(L)$. Triangles 
squares and circles denote the $N=4$, $N=3$ and $N=2$  data respectively.
Errors are not reported since they are smaller than the symbol sizes.
To render easier the comparison among the three sets of data, all 
the values of $D_R$ have been normalized to the best fit value 
of the constant $C$  (see tab.9 for the value of $C$). The three lines
correspond to the best fits obtained neglecting the $L=6$ derivative.

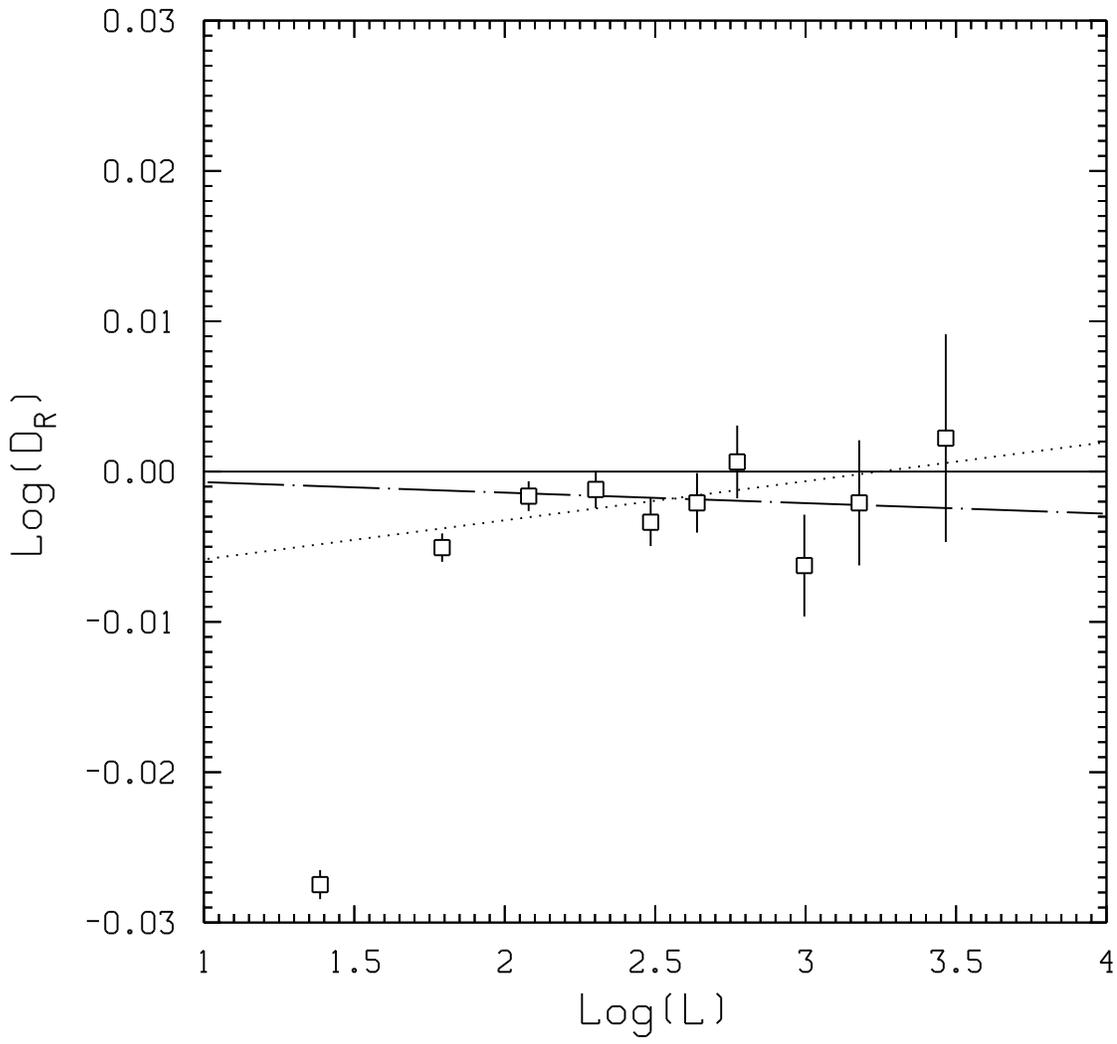
\item{\bf Fig.3}
The $N=3$ data only, plotted with a much higher resolution. The dotted line
corresponds to the best fit {\sl including } $L=6$, while the dashed line
corresponds to the $L=[8-32]$ fit. All the points 
 are normalized as in fig.2.

\end{description}

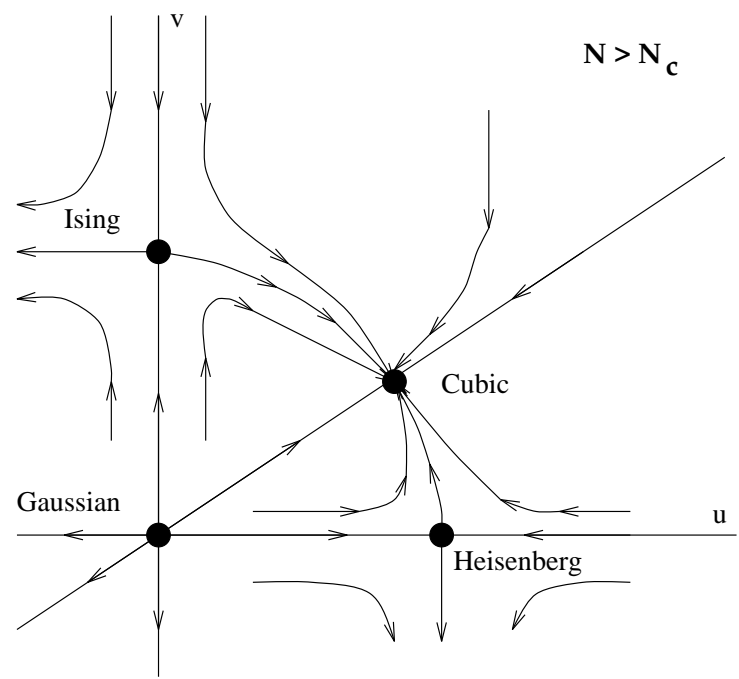
\end{document}